\documentclass[epjCONF,columns]{svjour}
\usepackage{graphics}
\usepackage[varg]{txfonts} 
\usepackage[latin1]{inputenc}
\session-title{2011 Hadron Collider Physics symposium (HCP-2011), Paris, France, November 14-18 2011}
\begin{document}
\title{Electroweak results at LHCb}
\author{Yasmine Amhis\inst{1}\fnmsep\thanks{\email{yasmine.amhis@epfl.ch}}  on behalf of the LHCb collaboration}
\institute{\'Ecole Polytechnique F\'ed\'erale de Lausanne, Switerland}

\abstract{$W$ and $Z$ production cross-sections have been
measured in proton-proton collisions at a center-of-mass energy of 7 TeV using the decays $W\rightarrow \mu \nu$, $Z\rightarrow \mu \mu$ and $Z\rightarrow \tau \tau$ collected by the LHCb detector.
For all the measurements at least one of the reconstructed muons has a transverse momentum, $p_T$, above 20 GeV/$c$ and a pseudorapidity, $\eta$, 
between 2 and 4.5. In the case of the $Z$, a di-muon invariant mass between 60 GeV$/c^2$ and 120 GeV$/c^2$ is required. Theoretical predictions, calculated at
next-to-next-to-leading order in QCD using recent parton distribution functions (PDFs), are found to be in agreement with the measurements.
} 
\maketitle
\section{Introduction}
\label{sec:intro}
\begin{figure}
\centering
\resizebox{0.60\columnwidth}{!}{%
\includegraphics{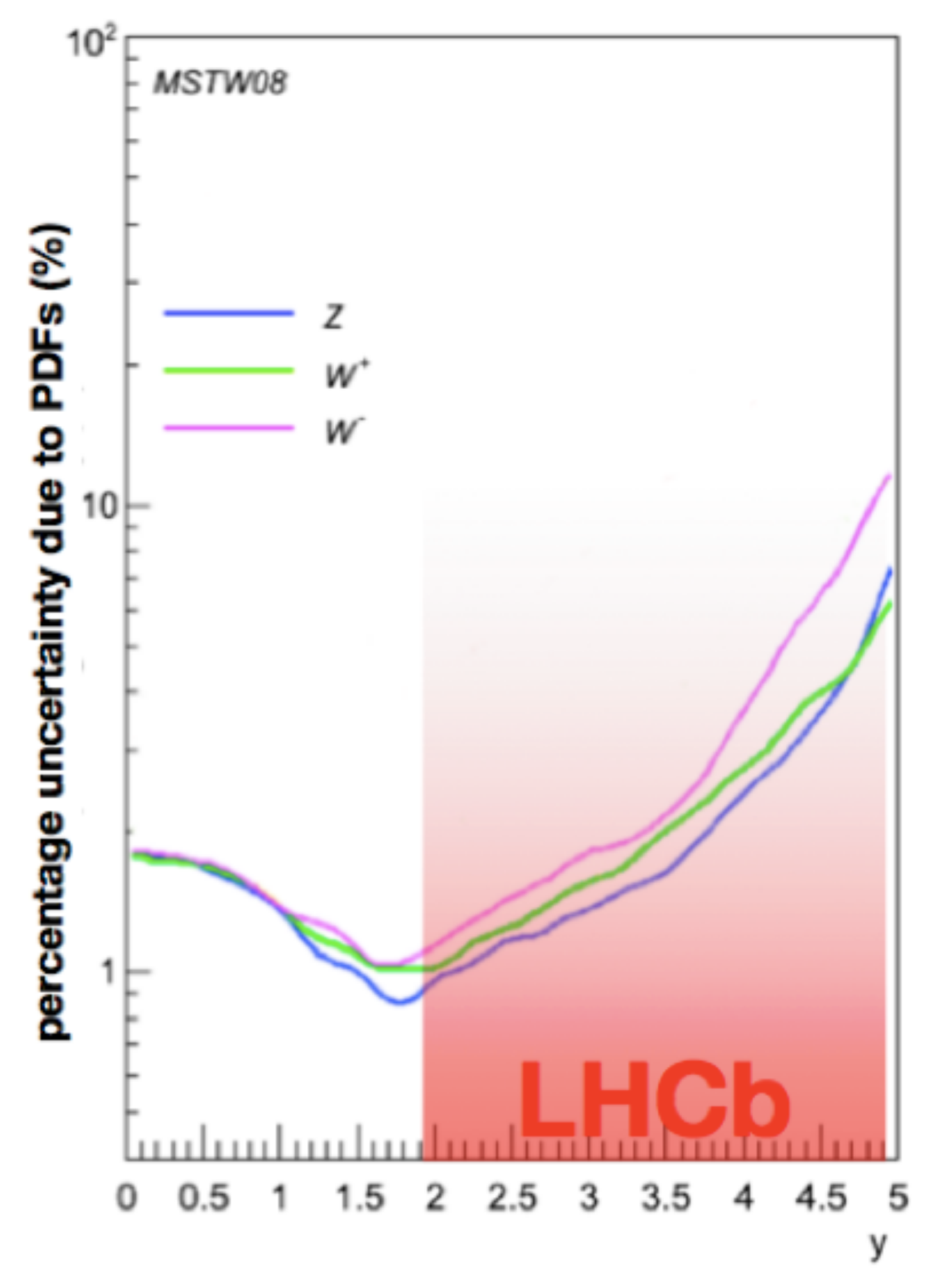} }
\caption{Percentage uncertainty due to PDFs as a function of rapidity for $Z$, $W^{\pm}$ bosons production cross-sections.}
\label{fig:PDFXSection}       
\end{figure}

\begin{figure}
\centering
\resizebox{0.60\columnwidth}{!}{%
\includegraphics{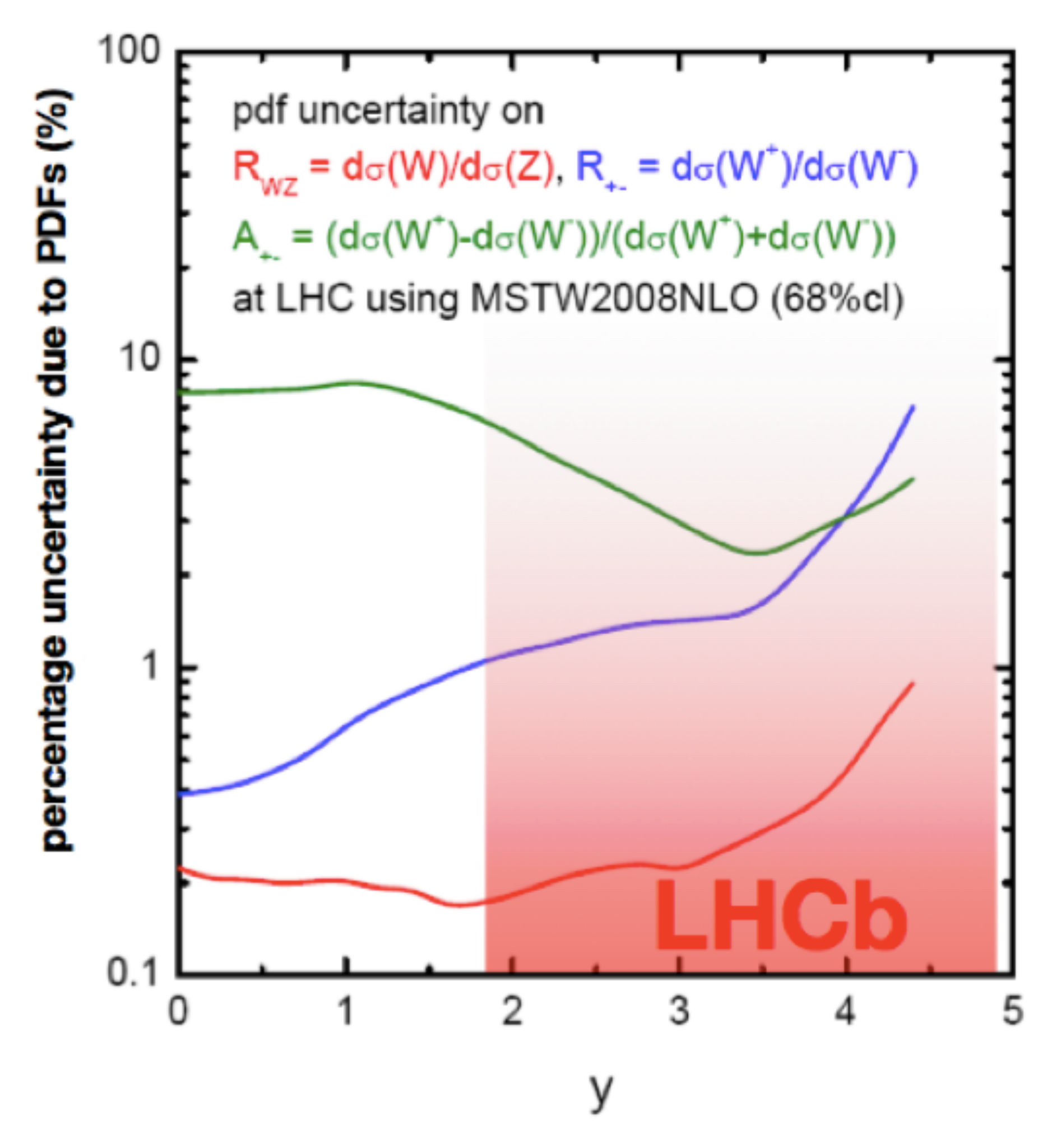} }
\caption{Percentage uncertainty due to PDFs  as a function of rapidity for $Z$, $W^{\pm}$ bosons ratios.}
\label{fig:PDFRatio}       
\end{figure}

Measurements of $W$ and $Z$ cross-sections in proton-proton collistions constitute an important test of the 
Standard Model.
While electroweak theory can currently describe the fundamental partonic processes of electroweak boson production at the LHC at NLO with an accuracy at the percent level, a large uncertainty on the theoretical predictions arise from the present knowledge of the proton PDFs. The accuracy strongly depends on the rapidity range as shown in Fig.~\ref{fig:PDFXSection}~and~\ref{fig:PDFRatio} : in kinematic regions where PDF uncertainties are low, precise measurements of electroweak bosons provide a stringent test of the Standard Model in a new energy regime; in regions where PDFs are less known, production studies can input valuable information to constrain new PDF fits.
The LHCb experiment, described in more detail in \cite{LHCbDetector}, at the CERN Large Hadron Collider (LHC) has been designed 
to study heavy flavour physics in the forward region ($1.9 < \eta < 4.9$). 
Measurements provided by LHCb can test the Standard Model and provide input to constrain the 
PDFs in both a region accessible only to LHCb 
($\eta > 2.5$) and a rapidity region which is also accessible to ATLAS and CMS ($2 < \eta < 2.5$). 
The following electroweak measurements at LHCb include the analysis of the $Z$ and $W$ boson 
production at $\sqrt{s} = 7$~TeV. In addition, the measurement of $Z$ boson production in decays to pairs of $\tau$ leptons is reported. 
The cross-section ratios $\sigma_W/\sigma_Z$ and $\sigma_W^+/\sigma_W^-$, and the $W$ charge asymmetry, $A_W$, are also presented. 
Further information can be found in Refs~\cite{lhcbconf039},\cite{lhcbconf041}.
\section{Event selection}
\subsection{Selection of $Z\rightarrow \mu\mu$ candidates}
\begin{figure}
\centering
\resizebox{0.80\columnwidth}{!}{%
\includegraphics{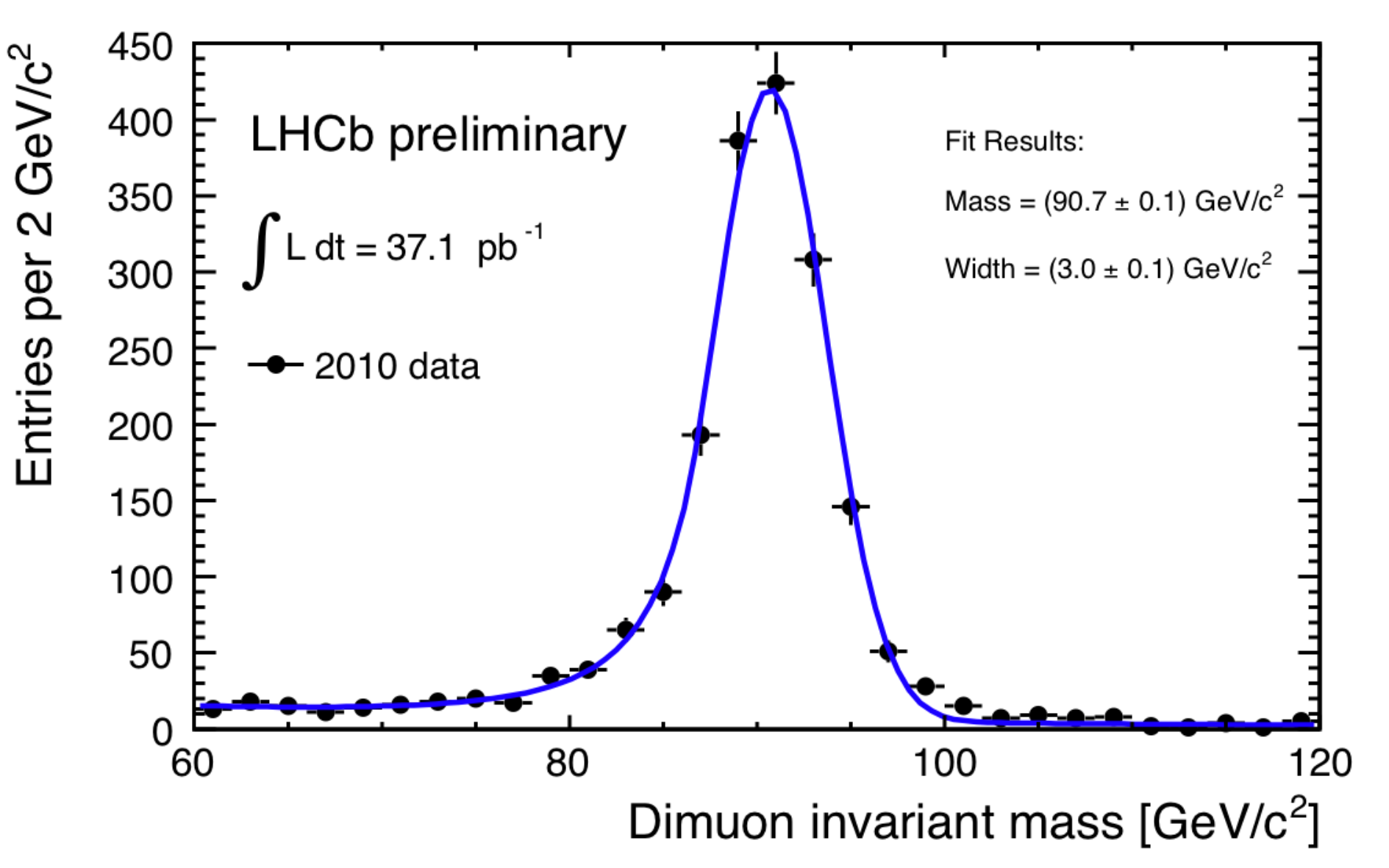} }
\caption{Dimuon invariant mass of $Z$ candidates. Data points are fitted with a Crystal Ball function (signal) 
and an exponential function (background).}
\label{fig:Z}       
\end{figure}

$Z$ candidates are selected requiring two well reconstructed muons with a transverse momentum, $p_T$, greater than
20 GeV/$c$ and lying in the pseudo-rapidity range between 2.0 and 4.5. 
The di-muon invariant mass distribution of such candidates is shown in Fig:~\ref{fig:Z}. 
To further identify $Z\rightarrow\mu\mu$ candidates, the invariant mass is required to be consistent 
with $Z$ production by imposing the mass constraint 60 GeV/${c^2}$ $ < m_{\mu\mu} < 120 $ GeV/${c^2}$. 
1966 candidates satisfy these criteria.
Several sources of background are considered: 
\begin{itemize}
\item [i] $Z\rightarrow\tau\tau$ decays where both taus decay to muons. Using simulation this contamination is 
estimated to be $(0.61\pm 0.04)$ events;
\item [ii] Heavy flavour background containing $b$ or $c$ quarks which decay semi-leptonically is estimated to be $(4.3 \pm 1.7)$ events;
\item [iii] Generic QCD events where pions or kaons either decay in flight or punch-through the detector to be falsely identified 
as muons. Using same sign di-muon combinations in data this contamination is expected to be $0 \pm 1$ events.
\end{itemize}
The total background estimate is found to be $(4.9\pm 2.0)$ events.
\begin{figure}
\centering
\resizebox{0.80\columnwidth}{!}{%
\includegraphics{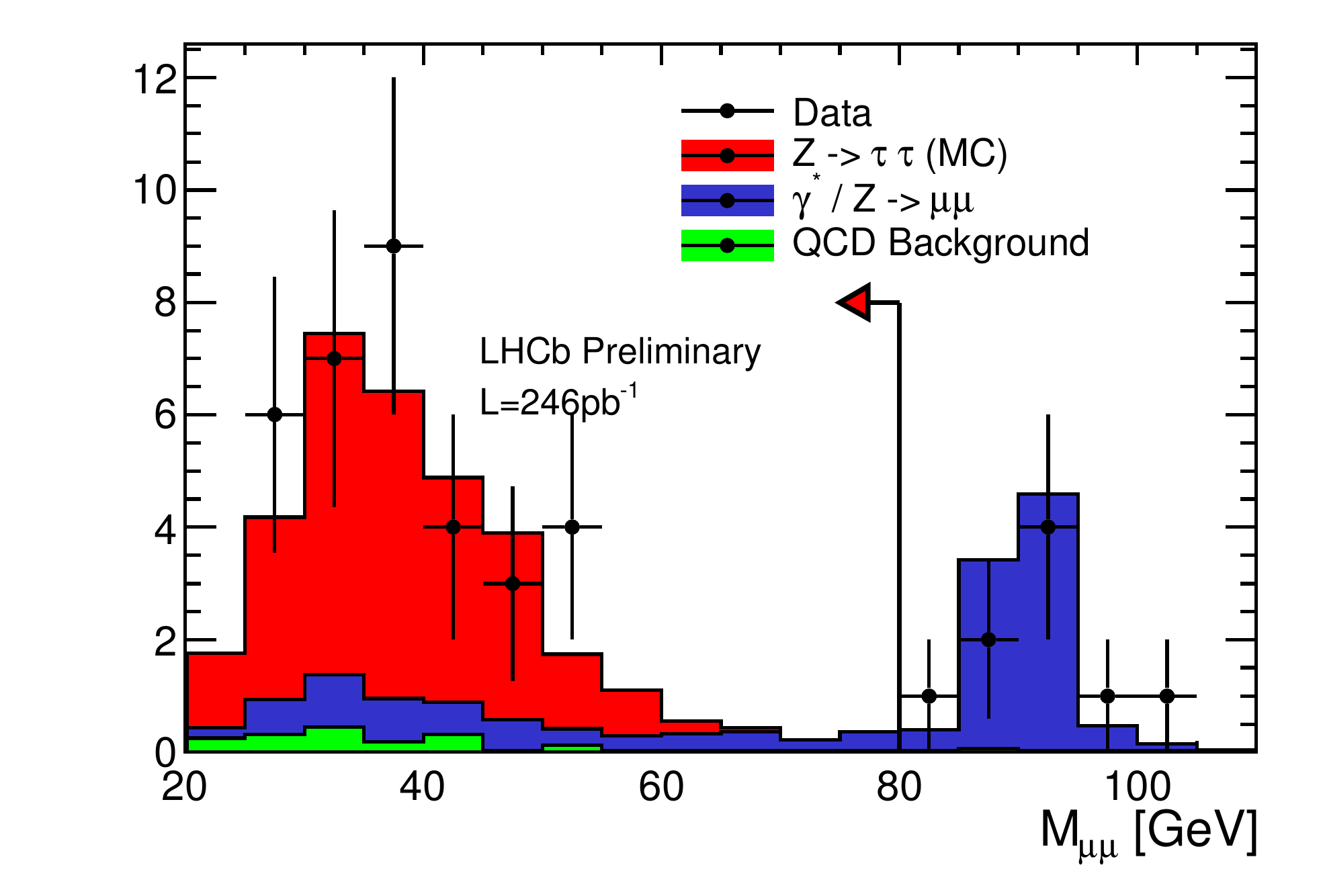} }
\caption{Invariant mass of the dileptons in the $\mu\mu$ channel. The points are data while the solid histograms show the estimated  signal (red) and background contributions (blue and green).}
\label{fig:Z2}       
\end{figure}
\begin{figure}
\centering
\resizebox{0.80\columnwidth}{!}{%
\includegraphics{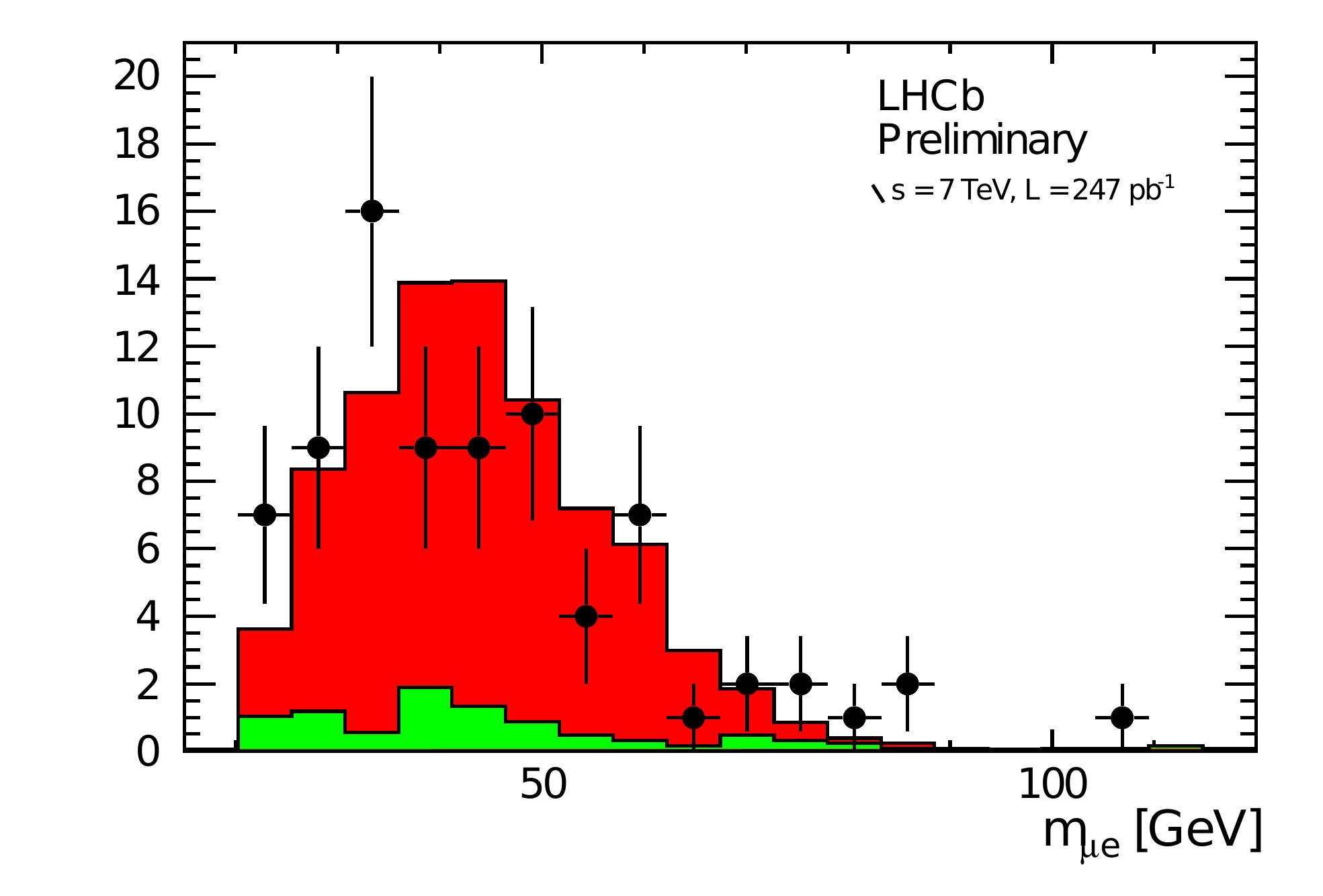} }
\caption{Invariant mass of the dileptons in the $e\mu$ channel. The points are data while the solid histograms show the estimated signal (red) and background contribution (green).}
\label{fig:Z1}       
\end{figure}
\subsection{Selection of $Z\rightarrow \tau\tau$ candidates}
Two different final states are considered to measure the $Z\rightarrow \tau\tau$ production cross-sections. 
In the first topology both tau leptons decay to muons and neutrinos, while for the second one, one tau decays into a muon and  neutrinos 
and the other decays to an electron and neutrino. For both channels, a muon with a $p_T$ greater than 20 GeV/$c$ and a pseudo-rapidity 
between 2 and 4.5 is required. Since both channels contain a high $p_T$ muon. 
For both topologies the invariant masses are shown in Fig:~\ref{fig:Z2},~\ref{fig:Z1}.
The following sources of background have been considered: 
\begin{itemize}
\item[i] QCD, estimated from same-sign and non-isolated data as $(9.5 \pm 3.0)$ and  $(1.6 \pm 1.3)$ events for electron-muon and muon-muon channels respectively; 
\item[ii] Electroweak, di-boson, and top backgrounds are estimated using simulated data. Drell-Yan backgrounds in the di-muon channel are estimated from data,
by fitting the di-muon invariant mass above 80 GeV$/c^2$ to the expected shape from this contributions. We find  $(3.0 \pm 1.2)$ and  $(5.5 \pm 1.8)$
events for the electron-muon and muon-muon channels respectively. 
\end{itemize}
\subsection{Selection of $W\rightarrow \mu\nu$ candidates}
The signature for $W$ boson decays is characterised by a single isolated high transverse momentum lepton and minimal other activity in the event. 
The background contamination is expected to be larger than in the case of $Z$ decays. 
To suppress the background additional cuts are imposed. 
Backgrounds from punch-through kaons and pions are suppressed by requiring that the sum of hadronic and electromagnetic calorimeter energies associated with the muon candidate, divided by the track momentum, is less than 0.04. Isolation is imposed by requiring that the summed p$_T$ of all tracks and identified photons inside a cone of radius  surrounding the muon, $R= \sqrt{\Delta\eta^2 + \Delta\phi^2} = 0.5$, should be less than 2 GeV. The unbiased impact parameter, $IP$, of the muon must be less than 40 $\mu$m to reduce backgrounds from heavy flavour production. Contributions from $\gamma^*/Z$ decay are suppressed by vetoing any other muon in the event with a $p_T$ above 5 GeV/$c$.
Using this selection 15608 $W^{+}$ candidates and 12301 $W^{-}$ candidates are retained.  
The following sources of backgrounds have been taken into account: 
\begin{itemize}
\item[i]  $Z\rightarrow \mu\mu$ where one of the muons goes outside of the LHCb acceptance; 
\item[ii]  $W\rightarrow \tau \mu_{\tau}$ where the tau decays leptonically to a muon and neutrinos;
\item[iii] $Z\rightarrow \tau\tau$ where the taus decay  leptonically to a muon and a neutrino;
\item[iv] Low mass Drell-Yan production of di-muon final states, where on muon is outside the LHCb acceptance; 
\item[v]  Events containing $b$ or $c$ quarks decaying semi-leptonically to a muon;
\item[vi] Generic QCD events where pions or kaons decay in flight;
\item[vii] Generic QCD events where pions and kaons punch-through. 
\end{itemize}
The signal yields are estimated by fitting the muon $p_T$ spectrum to shapes expected for signal and backgrounds in five bins of the muon pseudo-rapidity. 
The fit is performed for both charges and over all the $\eta$ bins simultaneously; only the normalisations of the signal and background (iv) are 
allowed to vary. The signal shapes are obtained using NLO simulation, using POWHEG~\cite{powheg} and  CTEQ6m~\cite{cteq}  parton density function set.
The shapes for backgrounds (i) to (iv) are determined using simulation (POWHEG for (i), Pythia~\cite{pythia} otherwise).
The normalisation of (i) is constrained to the observed rate of Z production, and (ii) to the observed rate of W production, taking branching ratios of $\tau$
to muons into account. The contributions of (iii) and (iv) are estimated from simulation. The shapes for backgrounds (v), (vi) and (vii) are found directly
from data by anti-cutting on selection variables to enrich these sources. The normalisations are fixed to the rates observed in data, with the exception of
decays in flight which are allowed to vary in the fit. The purity of the $W^{+}$ ($W^{-}$) sample is 80\% (78\%).
\begin{figure}
\centering
\resizebox{0.8\columnwidth}{!}{%
\includegraphics{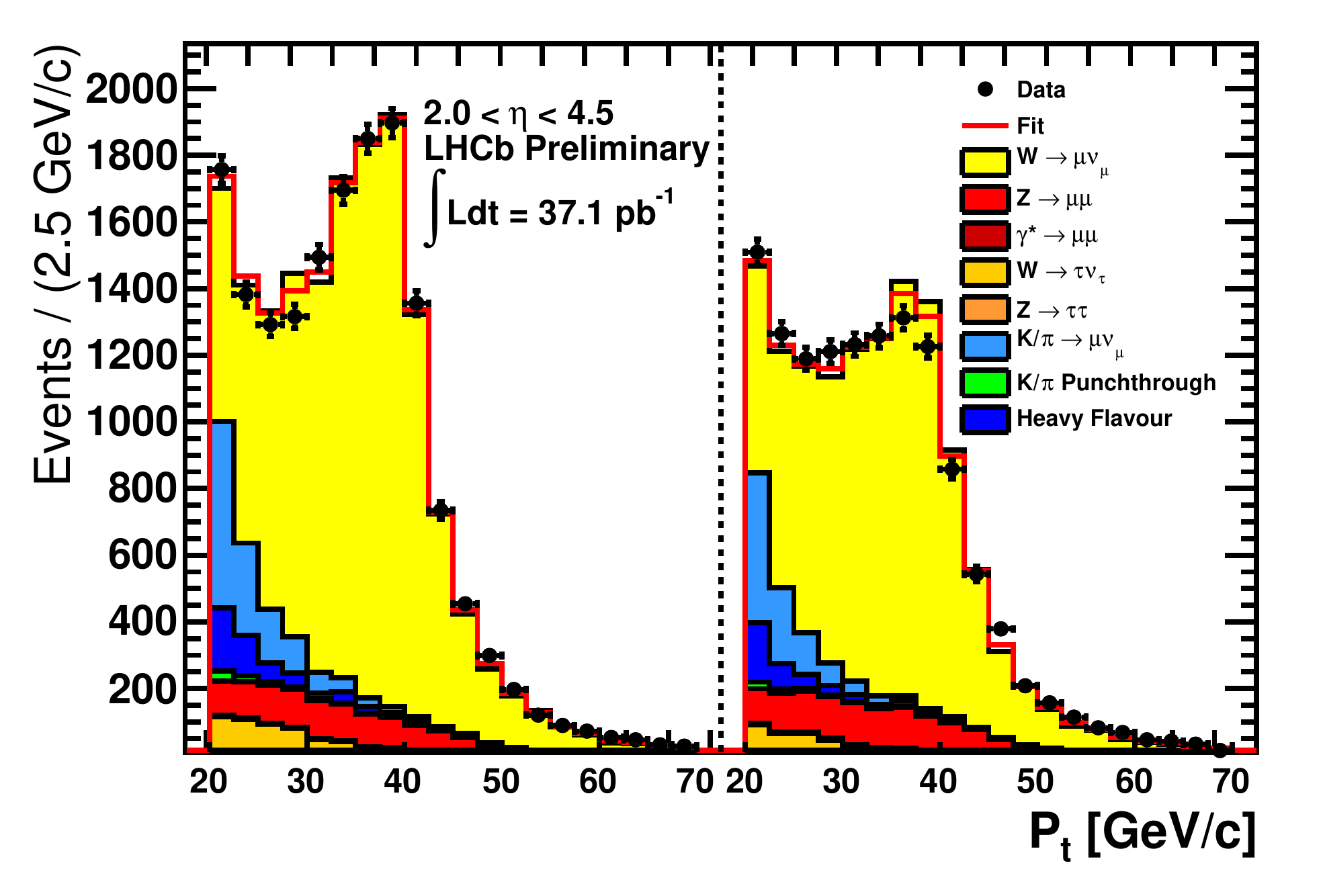} }
\caption{Distribution of muon $p_T$ for positive (left) and negative (right) charged muons. The data (points) are compared to the fitted contributions
from $W^-$ and $W^+$ (yellow), the $Z$ background in red (orange for $\tau\tau$ final states), ochre for $W$ decays to tau, the decay in flight background
in light blue, and heavy flavour (punch-through) contributions are shown in  blue (green).}
\label{fig:Wdistribution}       
\end{figure}
\section{Cross-section measurement}
The $Z$ production cross-section, $\sigma_Z$,  in a rapidity region, $\Delta y$, can be written 
\begin{equation}
\sigma_{Z\rightarrow \mu \mu}(\Delta y) = \frac{N_{tot}^Z - N_{background}^Z }{\epsilon_Z \times\cal{L}}, 
\end{equation}
where $N_{tot}^Z$ is the number of selected events, $N_{background}^Z$ is the estimated number of background events, $\cal{L}$ is the integrated 
luminosity corresponding to the analysed data sample,  and $\epsilon_Z$ is the total efficiency for selecting signal events. 
This efficiency is the product of an acceptance factor, $A^Z$, which is the proportion of events that should be reconstructed 
inside a given kinematic region; a trigger efficiency, $\epsilon_{trig}^Z$, which is the probability of triggering 
on such offline selected events; a tracking efficiency,  $\epsilon_{track}^Z$, which is the probability  of reconstructing 
both tracks of the $Z$; and finally a selection efficiency, $\epsilon_{sel}^Z$,  which is the probability of these leptons passing the final set of cuts. 
In the same way, we express the cross-section for $W$ production,  $\sigma_W$, in a muon pseudorapidity interval ,$\Delta \eta$, as 
\begin{equation}
\sigma_{W\rightarrow \mu \nu}(\Delta \eta) = \frac{N_{tot}^W - N_{background}^W }{\epsilon_W  \times\cal{L}}, 
\end{equation}
where now $\epsilon^W =  A^W\epsilon_{trig}^W \epsilon_{track}^W \epsilon_{muon}^W \epsilon^W_{sel}$, using the same notation as above, but with 
a $W$ superscript in the place of a $Z$ and where efficiency corresponds to one, and not two muons. 
One should note that all reported cross-sections are quoted within the fiducial region we measure, and not corrected to the 
full solid angle These measurements are corrected for final state radiations (FSR) to allow accurate comparison 
with theoretical predictions. The corrections are estimated using the HORACE generator~\cite{horace}. 
\subsection{Efficiency estimation}
Muon trigger, muon identification, and muon tracking efficiencies are determined from data-driven methods. 
A tag-and-probe technique is adopted using  $Z\rightarrow \mu\mu$ events, where one muon, used as a tag, is 
well identified and fired the single muon trigger, and the other muon or track candidate, used as probe, 
is tested to see how often it also fired the muon trigger, is identified as a muon, or has a track 
associated to it. These efficiencies are measured as a function of the lepton pseudo-rapidity. 
No dependence on the lepton charge is observed. Electron tracking efficiency is estimated from simulation,  and scaled by 
the difference in tracking efficiency observed between simulation and data from muons. 
The electron identification efficiency is determined from data using a tag-and-probe method using $Z\rightarrow ee$ events. 
\subsection{Systematic errors}
Systematic errors arise from the background models, the efficiency determination, the treatment of radiative correction applied 
to the cross-section measurement as well as the integrated luminosity error. 
The uncertainties on the background are due to the limited size of the simulated/real data. The systematic uncertainty on the efficiency is taken 
as the statistical precision on its determination from data, and covers the differences observed with respect to simulation.
The FSR correction uncertainty is given by the statistical uncertainty of the HORACE estimate. 
The luminosity is determined using two methods, a Van Der Meer scan and a beam gas method~\cite{lumi}. 
This systematic is dominated by the beam current uncertainty and is the dominating systematic for the $W$ and $Z$ 
production cross-section in the muon channels.  
In the case of $Z$ production with tau leptons in the final state, the uncertainty assigned on the efficiency determination is  the 
leading systematic. All systematic errors are summed in quadrature and reported in Tab:~\ref{tab:Wsystematic} and ~\ref{tab:Zsystematic}. 
\begin{table}
\caption{Contribution to the systematic error for the total $W$ production cross-section.  All quoted values are given in percentages. 
The luminosity is quoted separately.}
\label{tab:Wsystematic}
\begin{tabular}{lcc}
\noalign{\smallskip}\hline\noalign{\smallskip}
Source & $\Delta\sigma(W^+)(\%)$ &  $\Delta \sigma(W^-)(\%)$ \\
\noalign{\smallskip}\hline\noalign{\smallskip}
Background & 1.6 & 1.6 \\
Fit model & 1.9 & 1.7 \\ 
Efficiency & 2.5 & 2.3 \\ 
FSR correction & 0.2 & 0.2 \\
Total systematic error & 3.5 & 3.2 \\
Luminosity & 3.5 & 3.5 \\ 
\noalign{\smallskip}\hline\noalign{\smallskip}
\end{tabular}
\end{table}
\begin{table}
\caption{Contribution to the systematic error for the total $Z$ production cross-section. $e\mu$ and $\mu\mu$ refer to the final 
$Z\rightarrow\tau\tau$ production.  All quoted values are given in percentages.  The luminosity is quoted separately.}
\label{tab:Zsystematic}
\begin{tabular}{lccc}
\noalign{\smallskip}\hline\noalign{\smallskip}
Source& $\Delta \sigma(Z)$(\%)&$\Delta \sigma(e\mu)$(\%)&$\Delta \sigma(\mu\mu)$(\%)  \\
\noalign{\smallskip}\hline\noalign{\smallskip}
Background & 0.4 & 7 & 5  \\
Fit model & - & - &  - \\ 
Efficiency & 5.1 & 9 & 8  \\ 
Acceptance & - & 2 & \\
FSR correction & 0.3 & 0.2 & 0.2 \\
Total systematic error & 3.5 & 11 & 10 \\
Luminosity & 3.5 & 5.1  & 5.1\\ 
\noalign{\smallskip}\hline\noalign{\smallskip}
\end{tabular}
\end{table}
\begin{figure}
\center
\resizebox{1\columnwidth}{!}{%
\includegraphics{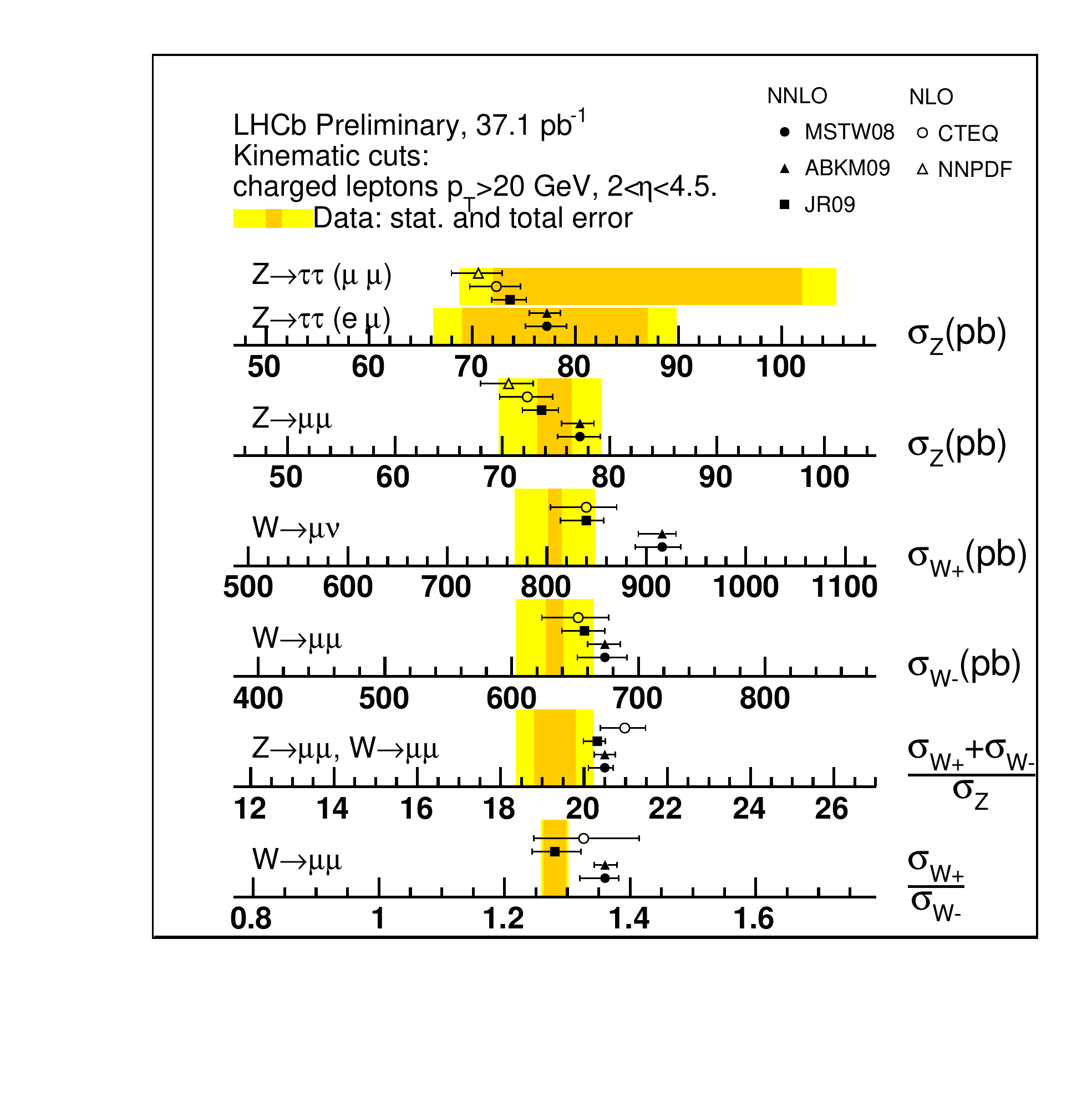} }
\caption{$Z$, $W^+$, $W^-$ cross-section measurements and ratios (coloured bands) compared to NLLO and NLO predictions (points), for 
the PDF set tested.}
\label{fig:Xsection}       
\end{figure}
\begin{figure}
\center
\resizebox{1\columnwidth}{!}{%
\includegraphics{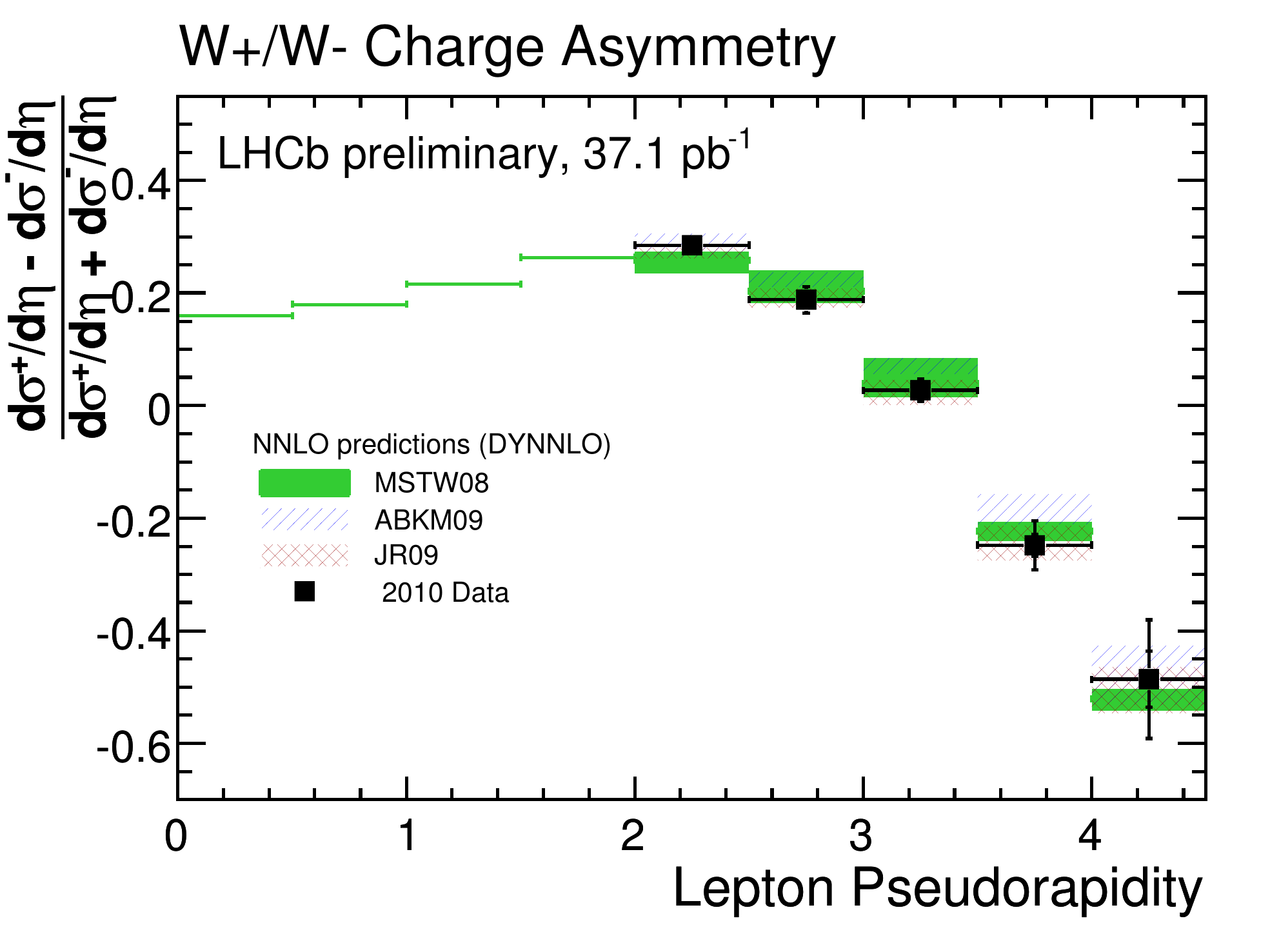} }
\caption{$W$ charge asymmetry in bins of muon pseudo-rapidity compared to the NNLO prediction. The shaded and hatched 
areas represent the uncertainty ariding from the PDF set tested.}
\label{fig:WA}       
\end{figure}
\section{Results}
The cross-section of $W$ and $Z$ production within the LHCb fiducial region are shown in Fig:~\ref{fig:Xsection} and 
the measured $W$ charge asymmetry is shown in Fig:~\ref{fig:WA}.  The measurements are shown as a coloured band on the summary and compared to various 
predictions based on next-to-leading-order and next-to-next-leading order calculations (NNLO) with a variety of PDF sets~\cite{Alekhin:2009ni},~\cite{jr},\cite{mstw}.
The $W$ charge asymmetry is shown as data points compared to NNLO prediction. 
All results are consistent with theoretical predictions. 
\section{Conclusions}
Measurement of the $W\rightarrow \mu\nu$ and $Z\rightarrow \mu\mu$ cross-sections and their ratios in proton-proon collision at $\sqrt{s} = 7 $ TeV 
have been performed using $(37.1\pm 1.3)$ pb$^{-1}$ of data collected by the LHCb detector.  Using $(240\pm$8)pb$^{-1}$of data collected both in 2010 and 2011, a measurement 
of $Z\rightarrow \tau\tau$ cross-section has also been made.  All measurements are in agreement with NNLO QCD predictions. Most of the errors are dominated 
by systematic uncertainties due to the determination of various reconstruction efficiencies from data. With larger statistics these 
systematic uncertainties will be reduced. Using the full 2011 dataset,  1 fb$^{-1}$, it is expected that improved measurements will place significant 
constraints on the proton PDFs.


\begin{thebibliography}{}
\bibitem{LHCbDetector} A. Augusto {\it et al.} JINST 3:S08005, 2008.
\bibitem{lhcbconf039}R.Aaij  {\it et al.}, LHCb Collaboration, LHCb-CONF-2011-039. 
\bibitem{lhcbconf041}R.Aaij  {\it et al.}, LHCb Collaboration, LHCb-CONF-2011-041. 
\bibitem{powheg} P.Nason, JHEP 0411 (2004)040, hep-ph/04091046; S. Frixione, P. Nason and C. Oleari, JHEP 0711 (2007) 070, arXiv:0709.2092; S. Alioli, P. Nason, C. Oleari and E. Re, JHEP 1006 (2010) 043, arXiv:1002.2581.
\bibitem{cteq}P. M. Nadolsky, H. -L. Lai, Q. -H. Cao, J. Huston, J. Pumplin, D. Stump, W. -K. Tung, C. -P. Yuan, Phys. Rev. D78 (2008) 013004. [arXiv:0802.0007 [hep-ph]].
\bibitem{pythia}T. Sjostrand {\it et al.}, Computer Phys. Commun. 135 238, 2001.
\bibitem{horace} C. M. Carloni Calame, G. Montagna, O. Nicrosini, M. Treccani, Phys. Rev. D69 (2004) 037301.
[hep-ph/0303102].
\bibitem{lumi} R.Aaij {\it et al.}, LHCb Collaboration 2012 JINST 7 P01010
\bibitem{Alekhin:2009ni}
 S.~Alekhin, J.~Blumlein, S.~Klein and S.~Moch, Phys.\ Rev.\ D {\bf 81} (2010) 014032 [arXiv:0908.2766 [hep-ph]].
\bibitem{jr}
 P. Jimenez-Delgado, E. Reya, Phys.\ Rev.\ {\bf D79} 074023 (2009), [arXiv:0810.4274 [hep-ph]]
\bibitem{mstw}
A. D. Martin, W. J. Stirling, R. S. Thorne and G. Watt, Eur.\ Phys.\ J.\ {\bf C63} 189 (2009), [arXiv:0901.0002 [hep-ph]]


\end{thebibliography}
\end{document}